\documentclass[]{aa501}
\usepackage{graphicx,psfig}
\newcommand{\ignore}[1]{}
\def\emu{{< \hspace{-4pt} \mu_e \hspace{-4pt}>}}
\def\mincir{\ \raise -2.truept\hbox{\rlap{\hbox{$\sim$}}\raise5.truept
\hbox{$<$}\ }}
\def\ea{{\it et. al.}}
\begin{document}
%
   \title{The fundamental plane of radio galaxies\thanks{Based on
   observations collected at European Southern Observatory, La Silla,
   Chile.}}


   \author{D. Bettoni \inst{1}, R. Falomo \inst{1}, G. Fasano  \inst{1},
F. Govoni \inst{2}, \\
M. Salvo \inst{1} and R. Scarpa \inst{3}
          }

   \offprints{D. Bettoni}

   \institute{$^{(1)}$Osservatorio Astronomico di Padova,
              Vicolo Osservatorio 5 35122 Padova, Italy\\
              \email{bettoni@pd.astro.it, falomo@pd.astro.it,
fasano@pd.astro.it, salvo@pd.astro.it} \\
$^{(2)}$Istituto di Radioastronomia di Bologna and Dipartimento di Astronomia,
Universit\'a di Bologna
\email{fgovoni@ira.bo.cnr.it} \\
$^{(3)}$European Southern Observatory
\email{rscarpa@eso.org} 
             }

   \date{Received; accepted }
\authorrunning{Bettoni et al.}


\abstract{
We collected photometrical and dynamical data for 73 
low red-shift (z$<$0.2) Radio Galaxies (LzRG) in order to study their
Fundamental Plane (FP). For 22 sources we also present new velocity
dispersion data, that complement the photometric data given in our
previous study of LzRG (Govoni \ea~2000a). It is found that the FP of
LzRG is similar to the one defined by non-active elliptical galaxies,
with LzRG representing the brightest end of the population of early
type galaxies. Since the FP mainly reflects the virial equilibrium
condition, our result implies that the global properties of
early--type galaxies (defining the FP) are not influenced by the
presence of gas accretion in the central black hole. This is fully in
agreement with the recent results in black hole demography, showing
that virtually all luminous spheroidal galaxies host a massive black
hole and therefore may potentially become active.  We confirm and
extend to giant ellipticals the systematic increase of the
mass-to-light ratio with galaxy luminosity.
\keywords {galaxies, radio-galaxies }
}


   \maketitle

%

\section{Introduction}

It is well known that the global properties of early--type galaxies
are fairly well described through a three dimensional space of
observables which, besides the effective radius $r_e$ and the
corresponding average surface brightness $\emu$, involves the central
velocity dispersion $\sigma_c$ (Djorgovski \& Davis 1987, Dressler \ea
~1987). In fact, ellipticals and lenticulars are found to lie in a
surprisingly tight, linear region of this space (the so called
Fundamental Plane; hereafter FP) that has been shown to be close to
the plane defining the virial equilibrium condition over the
assumption of a rigorous homology among galaxies (Faber \ea~1989).

The systematic deviation of the FP from the one defined by the virial
equilibrium, has been interpreted as due to changes of the Mass to
Light ratio (M/L) (due to different stellar content or to differences
in dark matter distribution; Pahre \ea~1995, Mobasher \ea~1999),
as well as to the breakdown of the homology assumption (e.g. kinematic
anisotropy; Caon \ea~1993, Graham \ea~1996, Ciotti \ea~1996,
Busarello \ea~1997).  In this context it is therefore important to
determine whether the FP has universal validity among ellipticals. In
particular it is not yet clear if active and non-active galaxies do
follow the same FP, as it would be expected in view of the recent
results on Black Holes (BH) demography in galaxies (Gebhardt \ea~2000,
Ferrarese \& Merritt 2000).

A relevant case is that of low redshift extragalactic radio sources
that are associated with massive early--type galaxies. A number of
optical studies have been carried out in the last decade to
investigate the relationship between radio emission properties and the
characteristics of the host galaxies (see e.g. Govoni \ea~2000a and
references therein).  Comparison of the optical properties of radio
galaxies with those of normal (non radio) ellipticals have shown that
the probability to exhibit radio emission increases with the optical
luminosity of the galaxies (Scarpa \& Urry 2001) and also that the
radio morphology (typically classes FR I and FR II) may depend on the
absolute magnitude of the host galaxy (Owen \& Laing 1989; Owen \&
White 1991). This could be related to the different BH mass
(proportional to the galaxy luminosity) and to the jet power
(Ghisellini \& Celotti 2001).

As a whole, these studies have shown that the optical 
morphological/structural and photometrical properties of
radio and non-radio ellipticals are remarkably similar, suggesting 
that all ellipticals may go through a phase of nuclear activity
lasting for a small fraction of the total life of the galaxy.  

Few previous observations were devoted to investigation of the
dynamical properties of Radio Galaxies (RG). Apart a very small number
(11) of radio sources included in the original work on the FP of
ellipticals by Faber \ea~(1989) (hereafter FA89), only Smith,
Heckman \& Illingworth~(1990; hereafter SHI90) performed a systematic
study on the stellar dynamics of Powerful Radio Galaxies (PRG). They
compared $\emu$, $r_e$ and $\sigma_c$ measurements for a compilation
of PRG with the distribution in the FP of the bright
ellipticals studied by FA89, concluding that the FP of PRG is
consistent with that of normal ellipticals. However they also found
evidence for smaller than normal velocity dispersions and significant
rotational support in galaxies with marked morphological peculiarity.

In this paper we present a much deeper investigation of the FP of
radio galaxies, based on a sample of 73 radio galaxies.  The data are
in part new (22 objects) and in part collected from the literature (51
objects). Most of literature values were derived from the Hypercat
database (Prugniel \& Maubon 2000). All values of $\emu$, $r_e$, and
$\sigma_c$ have been processed in order to make them homogeneous to a
common standard (see Sec 4.2). Throughout the paper we assume
$H_{\circ}$=50$h^{-1}$km $s^{-1}$Mpc$^{-1}$ and $q_0$=0.

\section{Observations and data analysis}

We obtained medium resolution optical spectra of RG selected from the
brightest ($m_R<$15) objects in the sample of 79 LzRG previously
imaged, in the R filter, by us (Fasano \ea~1996, Govoni
\ea~2000a,b). These observations were aimed at deriving the velocity
dispersion from stellar absorption lines. In Table~\ref{LOG} we give
the list of the objects observed together with exposure times and
position angles used for the observations.

Optical spectroscopy was obtained in March/April 1998 and November
1998 with the ESO/Danish~1.52~m telescope at La Silla, using the
Danish Faint Object Spectrograph and Camera (DFOSC). We used a CCD
Loral/Lesser, with 2052 $\times$ 2052 pixels combined with an Echelle
Grism of 316 grooves/mm yielding a velocity resolution (FWHM) of 71
km$s^{-1}$ (for a slit 1\arcsec\ wide) in the range
$\lambda\lambda$=4800-5800 \AA. A long slit, 2.0\arcsec\ wide and
centered on the galaxy, was oriented along the apparent major axis of
the radio galaxy. With this configuration we reach a velocity
dispersion resolution $\Delta\sigma$=60 km$s^{-1}$; the scale
perpendicular to the dispersion is 0.39\arcsec/pixel. Template
reference spectra of standard stars of spectral type from G8-III to
K1-III, with low rotational velocity (V$\times$$\sin{(i)}<$17
km$s^{-1}$), were secured at the beginning and at the end of each
night. During the observations the seeing ranged between 1\arcsec~and
1.5\arcsec.

\begin{table}
\caption[]{Log of the Observations}
\label{LOG}
\begin{tabular}{lccccr}
\hline
{\rm Object} & $\alpha$ & $\delta$ & z & {\rm Exptime} & {\rm $PA^{a}$} \\
& (2000) & (2000) & & sec & \\
\hline
\hline
0055-016  & 00 57 35 & -01 23 28 & 0.045 & 4800  &  90   \\
0131-360  & 01 33 58 & -36 29 35 & 0.030 & 3600  &  90    \\
0257-398  & 02 59 27 & -39 40 37 & 0.066  & 4800  & 120     \\
0312-343  & 03 14 33 & -34 07 40 & 0.067 & 3600  & 145   \\
0325+024  & 03 27 54 & +02 33 42 & 0.030 & 3600  & 155      \\
0449-175  & 04 51 21 & -17 30 14 & 0.031 & 3600  & 180      \\
0546-329  & 05 58 27 & -32 58 37 & 0.037 & 3000  & 180      \\
0548-317  & 05 50 49 & -31 44 26 & 0.034 & 3600  & 180    \\
0718-340  & 07 20 47 & -34 07 05 & 0.029 & 7200  &  75    \\
0915-118  & 09 18 06 & -12 05 43 & 0.054 & 4800  &  43   \\
0940-304  & 09 42 23 & -30 44 11 & 0.038 & 7200  &  87  \\
1043-290  & 10 46 10 & -29 21 10 & 0.060 & 7200  & 180    \\
1107-372  & 11 09 57 & -37 32 17 & 0.010 & 3600  &  32    \\
1123-351  & 11 25 53 & -35 23 40 & 0.032 & 6000  &  85    \\
1251-122  & 12 54 40 & -29 13 39 & 0.015 & 3600  &  17    \\
1333-337  & 13 36 39 & -33 57 56 & 0.013  & 3600  &  60    \\
1400-337  & 14 03 39 & -33 58 43 & 0.014  & 3600  &  84    \\
1404-267  & 14 07 29 & -27 01 02 & 0.022  & 4800  & 105   \\
1514+072  & 15 16 45 & +07 01 17 & 0.035  & 4800  &   5    \\
1521-300  & 15 24 33 & -30 12 20 & 0.020  & 5400  &  60   \\
2236-176  & 22 39 11 & -17 20 28 & 0.070  & 4800  & 124    \\
2333-327  & 23 36 07 & -32 30 30 & 0.052  & 4800  & 173    \\
\hline
\end{tabular}
\begin{list}{}{}
\item[$^{\mathrm{a)}}$] Slit position angle in degrees from North toward East
\end{list}
\end{table}

Optical spectra were reduced using standard procedures available in
the IRAF package and includes bias subtraction, flat fielding, and
wavelength calibration. The accuracy of the latter procedure was
checked with measurements of the night sky $\lambda_0$= 5577.32 \AA~
emission line. In all cases a precision of $\mincir$20 km$s^{-1}$ was
reached. In order to increase the signal to noise and to match the
observed spatial resolution (assuming the mean seeing) with the plate
scale across dispersion, the spectra were rebinned over three pixels
perpendicular to the dispersion, obtaining an effective spatial
resolution of 1.17\arcsec/pixel. The observed spectral range includes
the $Mg_2$ band ($\lambda_0$= 5175.4 \AA), the E-band (5269 \AA) and
the FeI line (5335 \AA).

The systemic velocity, corrected to the Sun, and the velocity
dispersion $\sigma$ were determined using the Fourier Quotient method
(Sargent \ea~1977, Bertola \ea~1984). The spectra were first
normalized by subtracting the continuum, converted to a logarithmic
scale and then multiplied to a cosine bell function that apodizes 10\%
of the pixels at each end of the spectrum. This forces the ends of
the spectra smoothly to zero. Finally the Fourier Transform of the
galaxy spectra were divided by the Fourier Transform of a template
star whose spectra (late G and early K spectral type) matched that of
the galaxy. These spectra are used as templates of zero velocity
dispersion.

The best-fit stellar template then yields a profile of the velocity
dispersion $\sigma$ and the stellar velocity curve with relative
errors for the galaxy. The $r.m.s.$ of the determinations obtained
with different template stars turned out to be less than 20
km$s^{-1}$ for $\sigma$ and $\sim$10 km$s^{-1}$ for the systemic
radial velocity $V_r$. The average values of these determinations were
adopted as final values of $\sigma$ and $V_r$.

Since early-type galaxies exhibit some gradients in the radial
velocity and velocity dispersion (Davies \ea~1983, Fisher \ea~1995),
the derived `central' parameter $\sigma_c$ depends on the distances of
the galaxies and the size of the aperture used for the observation. In
order to compare our velocity dispersions with the data available in
the literature we applied aperture corrections according to the
procedure given by J\o rgensen \ea~(1995). The individual measurements
of $\sigma$ are therefore corrected to a circular aperture with a
metric diameter of 1.19$h^{-1}$ kpc , equivalent to 3.4\arcsec\ at the
distance of the Coma cluster.

\begin{figure}
\resizebox{\hsize}{!}{\includegraphics*{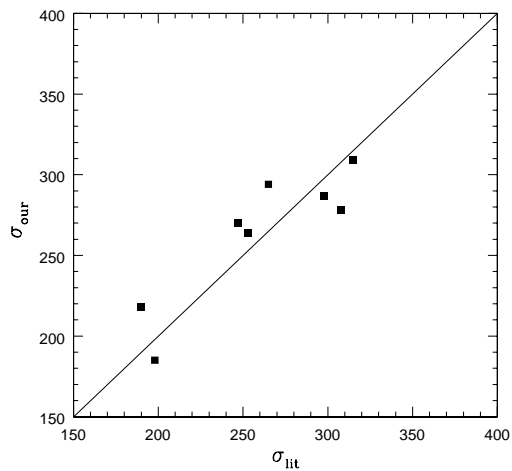}}
\vspace{-2cm}
\caption{Our velocity dispersion  $\sigma$ measurements compared with 
the literature data}
\label{Conf}
\end{figure}

For 8 of the 22 RGs observed previous determinations of the velocity
dispersion were found in the Hypercat database.  The comparison
between our and previous measurements of $\sigma$ is shown in Figure
\ref{Conf}. When more than one measurement of $\sigma$ is present in
the literature, we took the mean value quoted by Hypercat. The average
difference between our and Hypercat values of $\sigma$ is:

$<\sigma_{our}$-$\sigma_{Hyp}>$ = 4$\pm$6.4~km$s^{-1}$;\ \ $r.m.s.$ = 18~km$s^{-1}$

Additional notes on individual galaxies are given in the Appendix.

\begin{table}
\caption{Photometrical and kinematical Data for our sample of radio
galaxies}
\label{FP}
\begin{tabular}{lrclcr}
\hline
name & $\emu$ & log$r_e$ & log$\sigma$ & $\Delta$log$\sigma$ & $V_r~~~$ \\
& & kpc & km$s^{-1}$ & & km$s^{-1}$ \\
\hline
 0055-016 &  19.98  &  1.102  & 2.48 & 0.11 &  13593  \\    
 0131-360 &  20.16  &  1.145  & 2.40 & 0.17 &  8771    \\       
 0257-398 &  19.25  &  0.872  & 2.34 & 0.16 &  19814   \\     
 0312-343 &  21.25  &  1.352  & 2.41 & 0.07 &  20516   \\ 		
 0325+024 &  21.46  &  1.301  & 2.34 & 0.12 &  9875    \\ 	
 0449-175 &  20.50  &  1.163  & 2.20   & 0.15 &  9548    \\ 
 0546-329 &  20.47  &  1.267  & 2.59 & 0.34 &  11100   \\       
 0548-317 &  20.34  &  0.988  & 2.09 & 0.48 &  10352   \\       
 0718-340 &  19.70  &  1.043  & 2.52 & 0.50 &  9578    \\       
 0915-118 &  20.90  &  1.342  & 2.44 & 0.07 &  16171   \\       
 0940-304 &  19.72  &  0.973  & 2.59 & 0.20 &  12610   \\       
 1043-290 &  21.80  &  1.572  & 2.36  & 0.18 &  19527   \\       
 1107-372 &  19.66  &  1.093  & 2.47 & 0.09 &  3362    \\       
 1123-351 &  20.95  &  1.382  & 2.65 & 0.84 &  10632   \\       
 1258-321 &  20.40  &  1.237  & 2.42 & 0.12 &  5553    \\       
 1333-337 &  19.76  &  1.139  & 2.46 & 0.56 &  4122    \\       
 1400-337 &  20.70  &  1.408  & 2.49  & 0.20 &  4567    \\       
 1404-267 &  20.70  &  1.210  & 2.47 & 0.06 &  7187    \\       
 1514+072 &  21.89  &  1.620  & 2.43 & 0.12 &  11130   \\       
 1521-300 &  19.60  &  0.563  & 2.22 & 0.02 &  6579    \\       
 2236-176 &  21.33  &  1.492  & 2.39 & 0.15 &  22396   \\       
 2333-327 &  19.86  &  1.022  & 2.43 & 0.03 &  15707   \\       
\hline	    				       		   
\end{tabular}
\end{table}

Table~\ref{FP} reports our measurements of $\sigma_c$ and the estimated
uncertainities, together with the parameters $\emu$ and
$r_e$ , taken from Govoni \ea~(2000a), that are relevant
to construct the FP. The average surface brightness $\emu$ has been
evaluated from the formula: \\

$\emu$ = R$_T$ + 5log(r$_e^{\prime\prime}$) + 2.5log(2$\pi$), \\

\noindent
where R$_T$ is the total R apparent magnitude and
r$_e^{\prime\prime}$ is the effective radius. All these quantities
are corrected for the contribution of the point source (see Govoni et
al.~2000b), for cosmological dimming and for $K$--correction according
to Poggianti (1997). The distance/systemic velocities used to derive
$r_e$ and $\emu$ are relative to the Cosmic Microwave Background (CMB)
reference frame and come from the Lyon Meudon Extragalactic Database
(LEDA, Paturel \ea~1997).

\section{Extended sample of LzRG}

We have collected from the literature stellar velocity dispersion
measurements of LzRGs for which photometric and structural parameters
are also available. To make these data homogenous with our
measurements we processed them applying the same procedure adopted for
our subsample of LzRG (see above section and Govoni \ea~2000b).

We excluded from the analysis the radio galaxies belonging to dumbbell
systems or close galaxy pairs, as well as those having recession
velocity (corrected to CMB) less than 3000~km$s^{-1}$, since these
circumstances may induce significant uncertainties in the parameters
considered.

\subsection{Previous work on the Fundamental Plane of LzRGs}

An early investigation of the FP of LzRG was undertaken by FA89 in
their study of early-type galaxies. They give $\emu$ and $r_e$ (in the
B band), as well as $\sigma$ for 11 RGs. However 6 objects are dumbbell
systems or have small recession velocity and for this reason are not
considered in our analysis. For the remaining 5 we obtained $\emu$ in
the R band applying a color correction based on the integrated color
(B-R) obtained from Hypercat.

The first systematic study of the FP of RGs was performed by Smith,
Heckman \& Illingworth (1990, SHI90) who reported kinematical and
photometrical data for 20 powerful radio galaxies previously imaged in
the V band (Smith \& Heckman 1989, hereafter SH89), as well as for 16
more galaxies from Heckman \ea~(1985). Among the latter ones, only 7
are supplied with complete photometric information, whereas 5 objects,
out of the remaining 27 in the total sample, turn out to belong to
dumbbell systems or close pairs. This reduces the usable sample of
SHI90 to 22 objects available for FP analysis. In order to derive
$\emu$ consistently with our definitions (see above formula) we used
their isophotal magnitudes V$_{25}$ and the effective radii $r_e$
reported in SH89 (Table~8) and applied a correction of -0.16 (average
difference between m$_{tot}$ and m$_{24.5}$ in our study of RG; Govoni
\ea~2000b) to obtain total magnitudes. Then we corrected for the color
term (V-R) using the colors from SH89, when available, or assuming
(V-R)=0.55 in the other cases.

\subsection{Additional data for the Fundamental Plane of LzRG}

In order to improve the statistics of our analyses, we have searched
the literature for additional LzRG that have measurements of velocity
dispersion, $\emu$ and $r_e$. We have combined photometric data
of LzRGs from the studies of Ledlow \& Owen (1995, LO95) and
Gonzalez-Serrano \& Carballo (2000, GC00) whith the velocity
dispersions given by Wegner \ea~(1999, EFAR) or by Hypercat.

The cross identification betwen LO95 and EFAR added 17 new LzRGs to our
compilation, whereas that between GC00 and Hypercat adds 7 more
LzRGs. In both cases, in order to derive $\emu$ consistently with our
definitions, we used the magnitudes $m_{24.5}$ and the effective radii
$r_e$ given in the original papers and applied the same corrections
as for the data of SHI90.

In total, the sample of LzRG for which we collected data from the
literature consists of 51 objects. The relevant data for these
objects are given in Tables~\ref{FPlit} and \ref{FPlita}. These, together
with the sample for which we present new data, leads to a total 
of 73 LzRG, and represents the largest dataset of photometric and
spectroscopic data on Radio Galaxies till now.

\subsection{Comparison sample of non-radio  elliptical galaxies}

In order to compare the FP of LzRGs with that of normal galaxies, we
used the sample of radio-quiet early--type galaxies studied by J\o
rgensen \ea~(1996, JFK96). It is by far the richest, still homogeneous
sample for which both photometric and kinematic information is
available. Previous samples of early type galaxies used to describe
the FP (e.g. FA89) have a significantly larger scatter with respect to
the data of JFK96. Since the JFK96 sample consists of cluster
galaxies, it has also the advantage of the reliability of the distance
determination.

To compare our data on LzRG with the JFK96 data, we converted
their Gunn $r$ photometry to the R band, by means of the average
relation $R-r$=0.3, given in J\o rgensen~(1994).

\begin{table}
\caption[]{Radio Galaxies from literature}
\label{FPlit}
\begin{tabular}{llrcr}
\hline
name & log$\sigma$ & log$r_e$ & $\emu$ & $V_r~~~$ \\
& km$s^{-1}$ & kpc~ & &  km$s^{-1}$ \\
\hline
\multicolumn{5}{l}{Smith \ea~(1990)} \\ 
\hline
        3C29  &      2.318   &     1.180   &      20.29   &   13400    \\
        3C31  &      2.394   &     1.183   &      20.79   &    5007    \\
        3C33  &      2.362   &     1.066   &      20.43   &   17840    \\
        3C62  &      2.436   &     1.264   &      21.07   &   44370    \\
      3C76.1  &      2.301   &     0.784   &      19.28   &    9713    \\
        3C78  &      2.417   &     1.094   &      19.44   &    8634    \\
        3C84  &      2.391   &     1.234   &      20.37   &    5156    \\
        3C88  &      2.276   &     1.370   &      21.93   &    9054    \\
        3C89  &      2.398   &     1.610   &      22.51   &   41550    \\
        3C98  &      2.238   &     1.020   &      20.84   &    9174    \\
       3C120  &      2.301   &     1.146   &      20.51   &   10010    \\
       3C192  &      2.283   &     0.940   &      20.06   &   17930    \\
       3C196  &      2.322   &     1.464   &      21.23   &   59360    \\
       3C223  &      2.305   &     1.192   &      21.03   &   41010    \\
       3C293  &      2.267   &     0.876   &      18.99   &   13550      \\
       3C305  &      2.250   &     0.744   &      18.13   &   12470      \\
       3C338  &      2.462   &     1.562   &      21.48   &    9084      \\
       3C388  &      2.562   &     1.643   &      21.86   &   27220      \\
       3C444  &      2.190   &     1.763   &      22.62   &   45870    \\
       3C449  &      2.350   &     1.423   &      22.83   &    5126    \\
 PKS0634-206  &      2.290   &     0.878   &      18.70   &   16790    \\
 PKS2322-122  &      2.350   &     1.477   &      21.62   &    24610   \\
\hline
\multicolumn{5}{l}{Faber \ea~(1989)} \\
\hline
      NGC315  &      2.493   &    1.414    &   20.58  &    5126      \\
      NGC741  &      2.447   &    1.415    &   20.99  &    5546      \\
     NGC4839  &      2.387   &    1.311    &   21.41  &    6955      \\
     NGC7626  &      2.511   &    1.044    &   20.85  &    7615      \\
        3C40  &      2.233   &    0.640    &   18.18  &    5426      \\
\hline
\end{tabular}
\end{table}
\begin{table}
\caption[]{Radio Galaxies from literature, continue}
\label{FPlita}
\begin{tabular}{llrcr}
\hline
name & log$\sigma$ & log$r_e$ & $<\mu_e>$ & $V_r~~~$ \\
& & kpc & &  km/sec \\
\hline
\multicolumn{5}{l}{Ledlow \& Owen (1995)+ EFAR/Hypercat} \\
\hline
0039-095B&   2.447 & 0.362 & 17.69 & 16604 \\
0053-015 &   2.473 & 1.186  & 20.46 & 11459 \\
0053-016 &   2.396 &  0.866 & 19.57 & 12798 \\
0110+152 &   2.292 &  1.253 & 20.96 & 13137 \\
0112-000 &   2.401 & 0.799 & 19.63 & 13582 \\
0112+084 &   2.562 &  1.500 & 21.35 & 14740 \\
0147+360 &   2.384 & 0.601 & 18.57 &  5227 \\
0306+237 &   2.396 & 0.811 & 19.18 & 20122 \\
0431-133 &    2.430 &  1.382 & 21.02 &  9825 \\
0431-134 &   2.346 & 0.714 & 19.39 & 10393 \\
1510+076 &   2.526 & 0.455 & 18.45 & 12927 \\
1514+072 &   2.393 &  1.664 & 22.17 & 10308 \\
1520+087 &   2.342 &  1.497 & 21.85 & 10223 \\
1602+178A&   2.328 & 0.849 & 20.01 &  9350 \\
1610+296 &   2.508 & 0.814 & 18.76 &  9602 \\
2322+143a&    2.310 & 0.550 & 18.96 & 13381 \\
2335+267 &   2.538 &  1.349 & 20.71 &  9060 \\
\hline
\multicolumn{5}{l}{Gonzalez-Serrano \& Carballo (2000)+ Hypercat} \\
\hline
  NGC507 &2.517 &  1.005 & 19.72 & 4621 \\
  NGC708 &2.382 &  1.385 & 21.58 &  4498 \\
  gin116 &2.455 &  1.021 &  19.80 & 9791 \\
 NGC4869 &2.299 &  0.788 & 19.58 & 6895 \\
 NGC4874 &2.425 &  1.487 & 21.22 & 7428 \\
 NGC6086 &2.508 &  1.181 & 20.57 & 9534 \\
 NGC6137 & 2.47 &  1.506 & 20.66 & 9348 \\
\hline
\end{tabular}
\end{table}

\section{The Fundamental Plane of radiogalaxies}

In order to derive the parameters describing the FP: 

\begin{equation} 
\log r_e^{kpc} = \alpha\log \sigma + \beta\emu_R - \gamma , 
\end{equation} 

\noindent
we minimized the root square of the residuals perpendicular to the 
plane. This procedure is to be preferred with respect to the classical 
one (minimizing the root square of the residuals along a given axis) 
when there are measurement errors in all observed quantities. It is 
also slightly different from that used by JFK96, which minimizes the 
sum of the absolute residuals perpendicular to the plane. 

In Table~\ref{FIT} the FP coefficients $\alpha$, $\beta$
and $\gamma$ obtained using this fitting procedure for different 
subsamples of radio and non radio galaxies are reported.
Since the adopted fitting procedure does not provide
analytical form for the uncertainties of the parameters, we indicate
in Table~\ref{FIT} the 1$\sigma$ uncertainties computed according to
the classical formalism and assuming the values of the FP coefficients
given in the table.

Each sample in the table is indicated with a capital letter and the
fits obtained merging together two or more samples are indicated with
sequences of letters, according to the previous correspondence (for
instance JO means JFK96+This Work).

\begin{figure*}
\includegraphics[width=15cm]{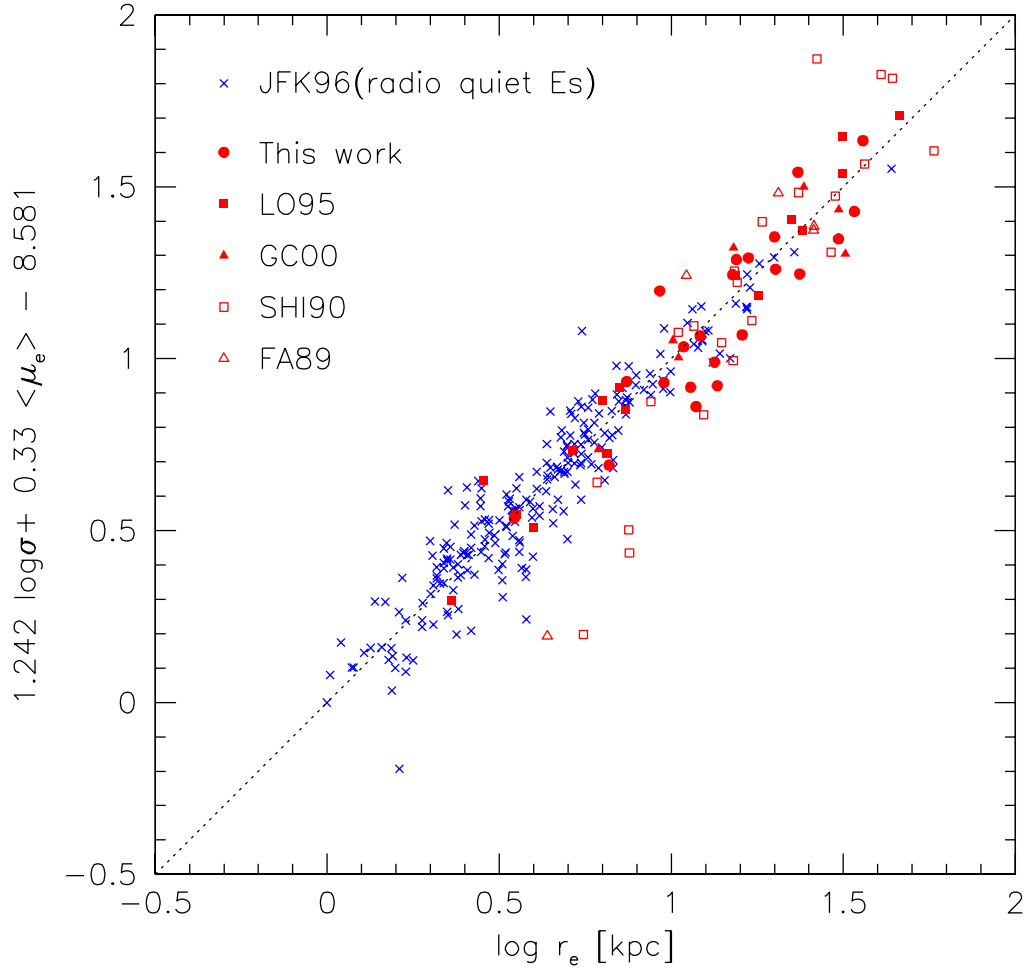}
\caption{The Fundamental Plane of various samples of LzRG (see legend and
Table~\ref{FIT}) and normal (non radio) ellipticals (JFK96), compared with
the best fit (dotted line) to the JFK96 data}
\label{FPRG}
\end{figure*}

We note from Table~\ref{FIT} that, in spite of the slightly different
fitting procedure, the FP coefficients we found for the JFK96 sample
of radio-quiet galaxies (our comparison sample) are indistinguishable
from those given in the original JFK96 paper (see their equation
[1]). In Figure~\ref{FPRG} we plot all the galaxies in the above described
samples, together with our FP fit of non radio ellipticals (JFK96).

From Table~\ref{FIT} the FP coefficients derived from our sample of 22
radio-galaxies turn out to be consistent (within 1$\sigma$ of
the estimated uncertainties) with those relative to the control sample
of non radio galaxies. The same happens for the 17 radio-galaxies in
the L095 sample. On the contrary, fitting the SHI90 sample of 22
radio-galaxies yields FP coefficients different (well beyond
3$\sigma$) from those relative to the comparison sample. This is likely
to be due to the presence of a few outliers in their galaxy sample, as
shown in Figure~\ref{FPRG}. Finally, given the small size of FA89 and
GC00 samples a reliable estimates of the FP can not be derived from
them. Nevertheless, we have reported in Table~\ref{FIT} the
coefficients resulting from the formal fit of these samples. 
In the case of the GC00 sample, they are consistent with those
relative to the control sample, whereas considerable differences
(again caused by the presence of an outlier) come in the case of the
FA89 sample.

In Table~\ref{FIT} we also report the fits obtained merging together
two or more galaxy samples. Among them, we assume as
representative of the global FP of elliptical galaxies, the fit:

\begin{equation} 
\log r_e^{kpc} = 1.27\log \sigma + 0.326\emu_R - 8.56  
\end{equation} 

\noindent
which includes the comparison sample (JFK96) and the radio samples
from this work(O), from LO95(L) and from GC00(G).  This fit allows to
extend the FP up to the bright end of the luminosity function of
early--type galaxies.

\begin{figure*}
\includegraphics[width=15cm]{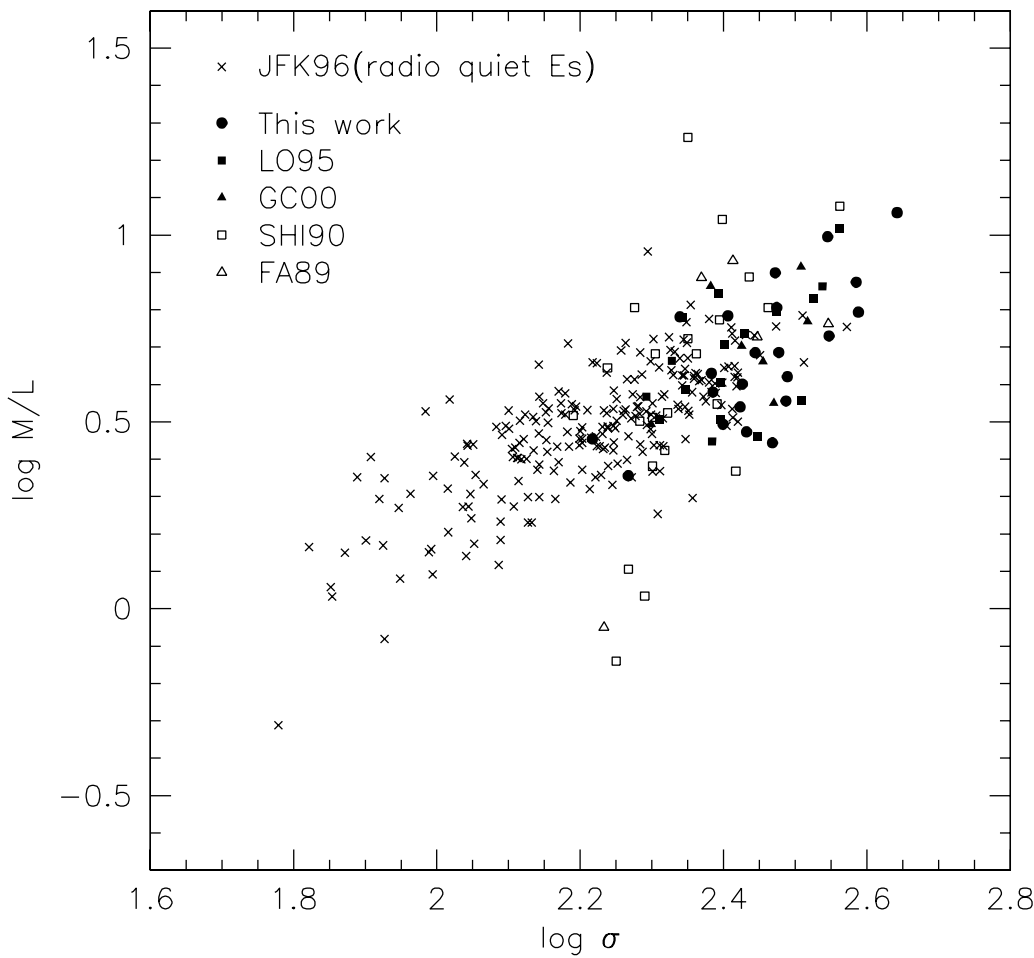}
\caption{The M/L ratio versus the velocity dispersion relationship 
of Radio Galaxies compared with that for early type galaxies (JFK96)}
\label{ML}
\end{figure*}

The M/L ratio is a function of the stellar population and dark matter
content of galaxies. In Figure~\ref{ML} we show the relation between
the Mass to Light ratio (M/L) and the velocity dispersion $\sigma$. We
derived the masses of our galaxies by using the relation
M=5$r_e\sigma^2$/G (Bender \ea~1992).  This figure indicates that
LzRGs follow the same relation found for normal elliptical
galaxies. The scatter of the M/L ratio as a function of log $\sigma$
is $\sim$0.2dex, similar to that found by JFK96.

\begin{table*}
\caption[]{The FP in R}
\label{FIT}
\begin{tabular}{lccccc}
\hline
Sample & $N_{gal}$ & $\alpha$ & $\beta$ & $\gamma$ & rms \\
\hline
J (JFK96) & 229 & 1.24$\pm$0.04 & 0.330$\pm$0.007 & 8.58$\pm$0.06 & 0.059 \\
O (This Work)& 22& 1.36$\pm$0.23& 0.295$\pm$0.031& 8.15$\pm$0.23 & 0.071 \\
L (LO95) & 17 & 1.24$\pm$0.23 & 0.311$\pm$0.015 & 8.21$\pm$0.18 & 0.047 \\
G (GC00) & 7 & 2.55$\pm$0.74 & 0.311$\pm$0.073 & 11.37$\pm$0.53 & 0.055 \\
S (SHI90)& 22 & 2.60$\pm$0.31 & 0.161$\pm$0.021 & 8.19$\pm$0.26 & 0.070 \\
F (FA89)& 5 & 2.31$\pm$0.48 & 0.072$\pm$0.043 & 5.86$\pm$0.50 & 0.030 \\
OLG & 46 & 1.58$\pm$0.16 & 0.310$\pm$ 0.015 & 8.99$\pm$0.15 & 0.063 \\
OLGSF & 74 & 1.92$\pm$0.15 & 0.256$\pm$0.013 & 8.69$\pm$0.15 & 0.082 \\
JOLG & 275 & 1.27$\pm$0.04 & 0.326$\pm$0.007 & 8.56$\pm$0.06 & 0.060 \\
JOLGSF & 303 & 1.35$\pm$0.04 &0.308$\pm$0.007 &8.40$\pm$0.06 &0.069 \\
\hline
\end{tabular}
\end{table*}
\section{Conclusions}

The main conclusion of this study is that the fundamental plane of
radio galaxies, as defined by our collection of data for 73 objects,
is consistent with the one defined by normal, non-radio elliptical
galaxies. Some recent results on BH demography (Ferrarese \& Merrit
2000, Gebhardt \ea~2000) strongly suggest that all galaxies may
host a massive BH in the nucleus, and that the BH mass is proportional
to the mass of the spheroidal component. This means that, virtually,
all ellipticals have the basic ingredient for becoming
active. Considering the small amount of gas (few solar masses per
year) necessary to sustain the radio emission of even the most
powerful radio sources, the activity could be triggered by extremely
modest alteration of the status of equilibrium in which galaxies
settled soon after they form. According to this view it is therefore
not surprising that radio and non radio ellipticals have
indistinguishable global properties, irrespective of their nuclear
activity. Here, we present one more result supporting this scenario.

Both metallicity effects and age variations have been suggested as the
main cause (Faber \ea~1995) of the strong dependence of M/L on
$\sigma$, that is on the total mass of the galaxy. This relation is
consistent with a progressive reddening of the stellar
population going from small to big galaxies (e.g., Prugniel \& Simien
1996), but most of dependence remains unexplained.  It looks like that
ellipticals pass from being baryon dominated to be dark matter
dominated with increasing luminosity, with a "fine tuning" required by
the small and basically constant dispersion around the FP.

\appendix\section{Notes on individual galaxies}

0055-016 - UGC 595 - 3C29 this galaxy belongs to the rich cluster A119
and is D26 in the Dressler (1980) list. Smith \ea~(1990) measured a
velocity dispersion of 199$\pm$18 km$s^{-1}$but recently Wegner
\ea~(1999) give a value of 253$\pm17$ km$s^{-1}$ in better agreement
with our measurements.

0131-360 - NGC 612 This galaxy show a strong dust lane along the apparent
major axis, our spectrum has been obtained perpendicularly to the dust
lane, i.e. along the minor axis of the optical galaxy, no rotation is
visible.

0915-118 - 3C218, Hydra A, A well studied galaxy showing strong
[OIII]($\lambda=5007$ \AA ) emission lines. Heckman \ea~(1985)
found $\sigma = 308\pm 38$ km$s^{-1}$, slightly higher than our measurement
$278\pm78$ km$s^{-1}$.

1107-372 - NGC 3557, We measure a velocity dispersion of 295$\pm$7
km$s^{-1}$ and a velocity gradient of $\sim$150 km$s^{-1}$ in the
central 5\arcsec\, in good agreement with data available in Hypercat.

1258-321 - ESO443-G024, this galaxy is in the cluster A3537. Dressler
\ea  (1991) measured a velocity dispersion of 279$\pm$27 km$s^{-1}$ in
good agreement with our data.

1333-337 - IC 4296, Our value of sigma is lower than the mean value
quoted by Hypercat (340 km$s^{-1}$). The agreement is better with the
measure of Franx \ea~(1989). From our spectrum, taken at
PA=60$^{\circ}$ we measure a rotation of $\sim$71 km$s^{-1}$ in agreement
with data taken at the same position angle by Saglia \ea~(1993).

1400-337 - NGC 5419, Our measure is in good agreement with the mean value
quoted by Hypercat. The spectrum, taken at PA=84$^{\circ}$, show a maximum
rotational velocity of $\sim$90 km$s^{-1}$ over 5\arcsec\ .

1514+072 - UGC9799, our measure of the velocity dispersion is very close
to the mean value of 247 km$s^{-1}$ quoted in Hypercat.

\begin{acknowledgements}
This research made use of Vizier service (Ochsenbein \ea~2000) and
of the NASA/IPAC Extragalactic Database (NED) which is operated by the
Jet Propulsion Laboratory, California Institute of Technology, under
contract with the National Aeronautics and Space Administration. We
have made use of the LEDA (http://leda.univ-lyon1.fr) and Hypercat
database. 
\end{acknowledgements}

\end{document}